\documentclass[10pt]{article}
\usepackage{geometry}
\usepackage{amsmath}
\usepackage{amssymb}
\usepackage{xcolor}
\definecolor{graycolor}{gray}{0.9} 
\usepackage{microtype}
\usepackage{setspace} 
\usepackage[utf8]{inputenc}
\usepackage[english]{babel}
\usepackage{times}
\usepackage{array}
\usepackage{soul}
\usepackage{cite}
\usepackage[numbers,sort&compress]{natbib}
\usepackage{natbib}

\usepackage{titlesec} 
\titleformat {\section} [block] {\raggedright \fontsize{10}{10}\selectfont\bfseries} {\thesection. \space} {0pt} {}
\titlespacing {\section} {0pt} {12pt} {6pt}
\titleformat {\subsection} [block] {\raggedright \fontsize{10}{10}\selectfont\itshape} {\thesubsection .\space} {0pt} {}
\titlespacing {\subsection} {0pt} {12pt} {6pt}
\titleformat {\subsubsection} [block] {\raggedright \fontsize{10}{10}\selectfont} {\thesubsubsection .\space} {0pt} {}
\titlespacing {\subsubsection} {0pt} {12pt} {6pt}
\titleformat {\paragraph} [block] {\raggedright \fontsize{10}{10}\selectfont} {} {0pt} {}
\titlespacing {\paragraph} {0pt} {12pt} {6pt}

\usepackage{cmap}
\usepackage[utf8]{inputenc}
\usepackage[T1]{fontenc}

\usepackage{array} \newcommand{\PreserveBackslash}[1]{\let\temp=\\#1\let\\=\temp}
\newcolumntype{C}[1]{>{\PreserveBackslash\centering}m{#1}}
\newcolumntype{R}[1]{>{\PreserveBackslash\raggedleft}m{#1}}
\newcolumntype{L}[1]{>{\PreserveBackslash\raggedright}m{#1}}
\usepackage{lineno}
\usepackage{tabularx}
\usepackage{colortbl}
\usepackage{graphicx}
\usepackage{float}
\usepackage[export]{adjustbox}
\usepackage{caption}
\usepackage{fancyhdr} 
\usepackage{lastpage}
\usepackage{layout}
\usepackage{setspace} 
\usepackage{enumitem}
\usepackage{booktabs}
\usepackage{arydshln}
\usepackage{multirow}
\usepackage{color}
\usepackage{hyperref} 
\hypersetup{
	colorlinks=true,
	linkcolor=blue,
	filecolor=blue,
	urlcolor=black,
	citecolor=cyan,
}

\fancypagestyle{firstpage}{
    \setlength{\headsep}{2.2cm}
    
    \setlength{\footskip}{1.5cm}
    \fancyhf{}% clear default for head and foot
    \lhead{\begin{table}[H]
        \centering
        \begin{tabular}{L{2.5cm}C{10cm}C{3.1cm}R{2cm}}
            \includegraphics[scale=0.035]{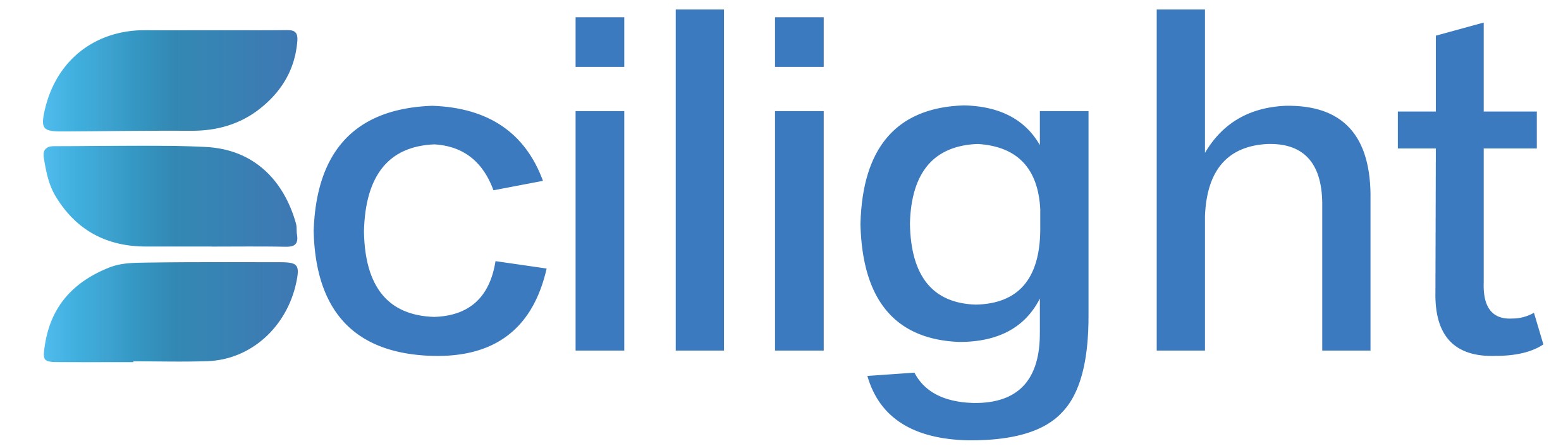} \vspace{-6pt}& \cellcolor{graycolor}\begin{tabular}[c]{@{}c@{}}\textit{International Journal of Gravitation and Theoretical Physics}\\ \href{https://www.sciltp.com/journals/ijgtp}{https://www.sciltp.com/journals/ijgtp}\end{tabular} & \includegraphics[scale=0.015]{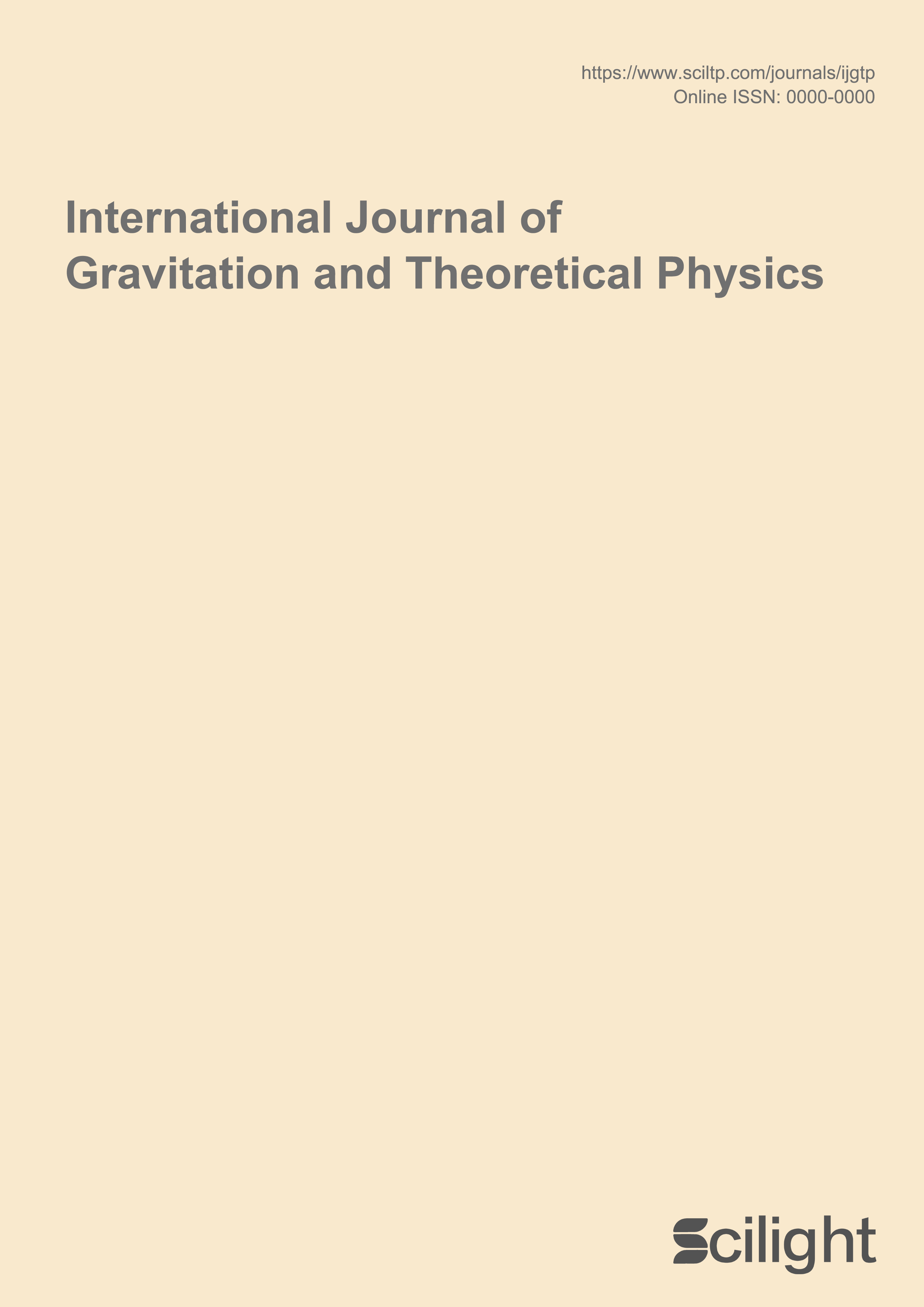} \vspace{-3pt}\\
        \end{tabular}
        \vspace{-22pt}
    \end{table}}
   
    \fancyfoot[C]{
        \vspace{-1.55cm}
        \begin{table}[H]
            \begin{minipage}[c]{0.15\columnwidth}
                \includegraphics[scale=0.5]{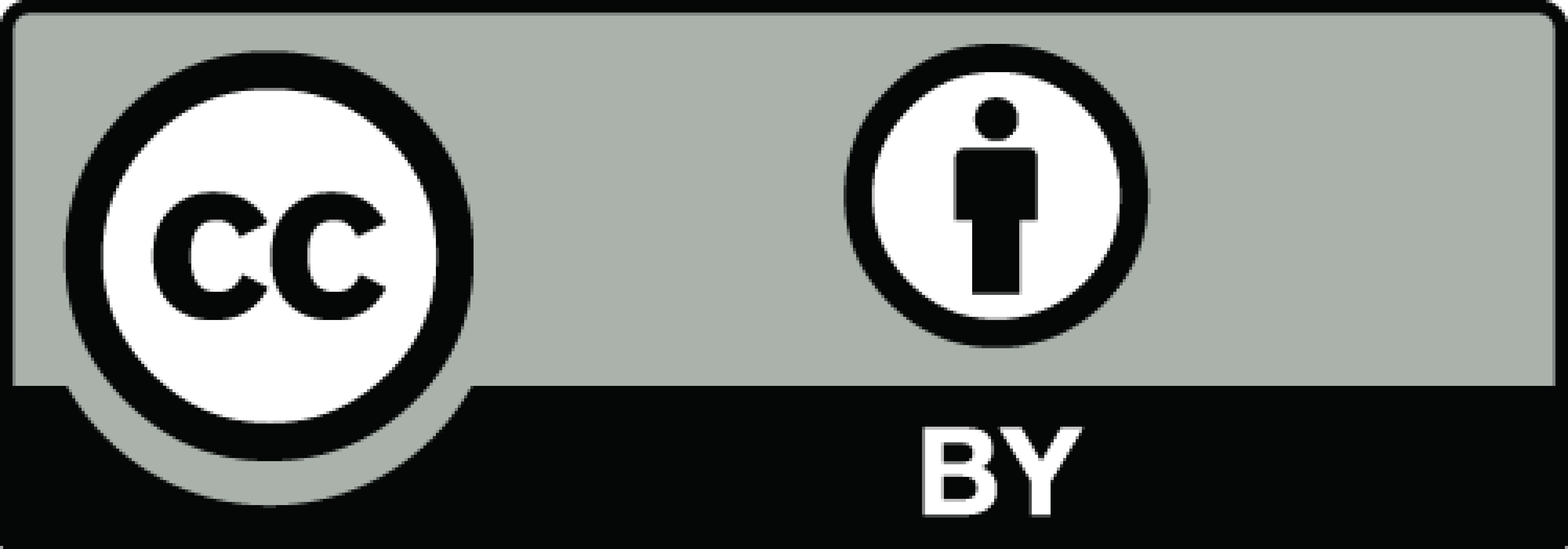} \vspace{1.1pt}
            \end{minipage}
            \hfill
            \begin{minipage}[c]{0.85\columnwidth}
                \scriptsize \textbf{Copyright:} © 2025 by the authors. This is an open access article under the terms and conditions of the Creative Commons Attribution
 (CC~BY) license (\href{https://creativecommons.org/licenses/by/4.0/}{https://creativecommons.org/licenses/by/4.0/}). \\ \textbf{Publisher’s Note:} Scilight stays neutral with regard to jurisdictional claims in published maps and institutional affiliations.
            \end{minipage}
    \end{table}}
    \vspace{-0.55cm}
}
\usepackage{color}

\begin{document}
\newgeometry{left=2.5cm, right=2.5cm, top=1.8cm, bottom=4cm}
	\thispagestyle{firstpage}
	\nolinenumbers
	{\noindent \textit{Article
}}
	\vspace{4pt} \\
	{\fontsize{18pt}{10pt}\textbf{Gravitational Quasinormal Modes and Grey-Body Factors of Bonanno–Reuter Regular Black Holes}  }
	\vspace{16pt} \\
	{\large S. V. Bolokhov \textsuperscript{*} and Milena Skvortsova}
	\vspace{6pt}
	 \begin{spacing}{0.9}
		{\noindent \small
			Peoples' Friendship University of Russia (RUDN University),  Institute of Gravitation and Cosmology,  
 6 Miklukho-Maklaya Street, 117198 Moscow, Russia \\
		    {*}  \parbox[t]{0.98\linewidth}{Correspondence: bolokhov-sv@rudn.ru} \\\\
		\footnotesize	\textbf{How To Cite}:  {Bolokhov, S.V.; Skvortsova, M. Gravitational Quasinormal Modes and Grey-Body Factors of Bonanno–Reuter Regular Black Holes. \emph{International Journal of Gravitation and Theoretical Physics} \textbf{2025}, \emph{Volume}(Issue), Page Number. \href{https://doi.org/10.xxxx/xxx}{https://doi.org/10.xxxx/xxx}}.}\\
	\end{spacing}

\begin{table}[H]
\noindent\rule[0.15\baselineskip]{\textwidth}{0.5pt} 
\begin{tabular}{lp{12cm}}  
 \small 
  \begin{tabular}[t]{@{}l@{}} 
  \footnotesize  Received: day month year \\
  \footnotesize  Revised: day month year \\
   \footnotesize Accepted: day month year \\
  \footnotesize  Published: day month year
  \end{tabular} &
  \textbf{Abstract:} We study gravitational perturbations of the Bonanno--Reuter quantum-corrected black hole arising in the asymptotic safety scenario, focusing on QNMs and grey-body factors. Assuming the RG parameter $\tilde{\omega}$ is fixed to its phenomenologically motivated value, we treat the interpolation parameter $\gamma$ as free and investigate how it modifies the black hole's response to axial gravitational perturbations. Quasinormal frequencies are computed using the sixth-order WKB method with Padé approximants, and their dependence on $\gamma$ and the black hole mass $M$ is analyzed. We find that the Schwarzschild limit is rapidly recovered for large $M$ or large $\gamma$, while significant deviations arise in the quantum regime. The accuracy of the WKB results is confirmed by time-domain integration of the wave equation. Comparison of grey-body factors computed via both the WKB method and the quasinormal mode correspondence are in a good concordance. Our findings indicate that quantum corrections can leave significant imprints in the ringdown and radiation spectra, while preserving consistency with classical results in the appropriate limit. \\
\\
  & 
\end{tabular}
\noindent\rule[0.15\baselineskip]{\textwidth}{0.5pt} 
\end{table}

\section{Introduction}

Quasinormal modes (QNMs) and grey-body factors are fundamental tools in the analysis of black hole spacetimes. The QNMs characterize the damped oscillations of perturbations around black holes and dominate the ringdown phase of gravitational wave signals~\cite{Kokkotas:1999bd,Berti:2009kk,Konoplya:2011qq,Bolokhov:2025uxz}. Their frequencies depend solely on the background geometry and the field content, making them powerful probes of gravitational dynamics and potential deviations from classical general relativity. On the other hand, grey-body factors describe the frequency-dependent transmission probabilities for quantum radiation escaping from black holes  \cite{Page:1976df,Kanti:2004nr}, modifying the ideal blackbody spectrum predicted by Hawking \cite{Hawking:1975vcx}. Together, these quantities encode both classical and semiclassical signatures of black holes and provide complementary windows into the structure of spacetime near the event horizon. In the context of quantum gravity proposals — such as asymptotic safety — it is of particular interest to understand how QNMs and grey-body factors are affected by quantum corrections, as these deviations may offer observational handles on quantum effects in strong-field regimes.

Understanding the interplay between quantum gravity effects and black hole physics is one of the most fundamental challenges in modern theoretical physics. While the classical theory of general relativity predicts the existence of singularities, quantum gravitational corrections are expected to modify the spacetime geometry at small scales, potentially resolving or softening such singularities. In the asymptotic safety program, the gravitational coupling becomes scale-dependent and runs with energy scale due to renormalization group (RG) effects \cite{Niedermaier:2006wt}. This leads to the concept of a “running Newton constant”, which, when incorporated into classical black hole solutions, results in quantum-improved metrics that differ significantly from their classical counterparts in the near-horizon and core regions.
\restoregeometry

In this context, Bonanno and Reuter~\cite{Bonanno:2000ep} proposed an effective Schwarzschild-like black hole metric derived by improving the classical Schwarzschild solution using the RG flow of the Newton constant. Their model captures essential features of asymptotic safety, including the emergence of a non-singular core for sub-Planckian masses and the existence of a minimal mass remnant that halts complete evaporation. While previous works have studied scalar and other test field perturbations in this background \cite{Konoplya:2022hll,Rincon:2020iwy,Liu:2012ee,Li:2013kkb}, the behavior of gravitational QNMs and grey-body factors has remained unexplored.

In this work, we extend the study of the Bonanno–Reuter black hole by computing the gravitational quasinormal mode spectrum and the associated grey-body factors. We employ both the sixth-order WKB method with Padé resummation and the time-domain integration method to extract the dominant modes and test their consistency. Furthermore, we analyze the grey-body factors using both the WKB technique and the recently proposed correspondence with QNMs. These results allow us to explore how the deviations from the classical Schwarzschild black hole depend on the quantum correction parameter $\gamma$ and the black hole mass $M$. Our findings demonstrate that the Schwarzschild limit is quickly recovered at large $M$ or large $\gamma$, while significant deviations appear in the quantum-dominated regime. The structure of the paper is as follows. In Section \ref{sec2}, we briefly review the Bonanno–Reuter black hole and its underlying theoretical framework. In Section \ref{sec3}, we derive the effective potential for axial gravitational perturbations. Section \ref{sec4} presents the essentials of the WKB method and time-domain approach as well as discusses features of the quasinormal spectrum. In Section \ref{sec5}, we analyze the grey-body factors, examining their parametric dependencies and mutual consistency. We conclude with a discussion of the implications of our results in Section \ref{sec6}.

\section{Renormalization Group Improved Black Hole}
\label{sec2}

The Bonanno–Reuter black hole arises from the renormalization group (RG) improvement of the classical Schwarzschild solution in the context of asymptotically safe gravity~\cite{Bonanno:2000ep}. In this approach, quantum gravitational corrections are encoded in a scale-dependent Newton constant $G(k)$, where $k$ is an energy scale that depends on the radial coordinate $r$. The effective Newton constant is modeled as
\begin{equation}
G(r) = \frac{G_0 r^3}{r^3 + \tilde{\omega} G_0 (r + \gamma G_0 M)},
\end{equation}
where $G_0$ is the infrared Newton constant, $\tilde{\omega} = 118/(15\pi)$ is fixed by matching the RG-improved potential to one-loop quantum corrections from effective field theory, and $\gamma > 0$ is an interpolation parameter determined by the choice of cutoff identification.

Substituting this $G(r)$ into the Schwarzschild lapse function $f(r) = 1 - 2 G(r) M / r$ yields the RG-improved metric:
\begin{equation}
ds^2 = -f(r)dt^2 + \frac{dr^2}{f(r)} + r^2d\Omega^2,
\end{equation}
with the lapse function explicitly given by
\begin{equation}
f(r) = 1 - \frac{2 G_0 M r^2}{r^3 + \tilde{\omega} G_0 (r + \gamma G_0 M)}.
\label{eq:lapse}
\end{equation}

This geometry exhibits several notable features:
\begin{itemize}[topsep=3pt,parsep=0pt,itemsep=0pt,leftmargin=*,labelsep=5.5mm,align=parleft]
\item For large $r$, the function $f(r)$ approaches the classical Schwarzschild behavior with leading corrections of order $1/r^3$.
\item For small $r$, the lapse function behaves as $f(r) \sim 1 - r^2 / r_0^2$, indicating a regular de Sitter core with effective cosmological constant $\Lambda_{\text{eff}} \sim 1/r_0^2$.
\item The metric may possess two, one, or no horizons depending on the mass $M$ and the parameter $\gamma$. There exists a critical mass $M_{\text{cr}}$ below which the black hole has no horizon, leading to a soliton-like remnant.
\end{itemize}

\begin{figure}[H]
\centering
\includegraphics[width=0.55\textwidth]{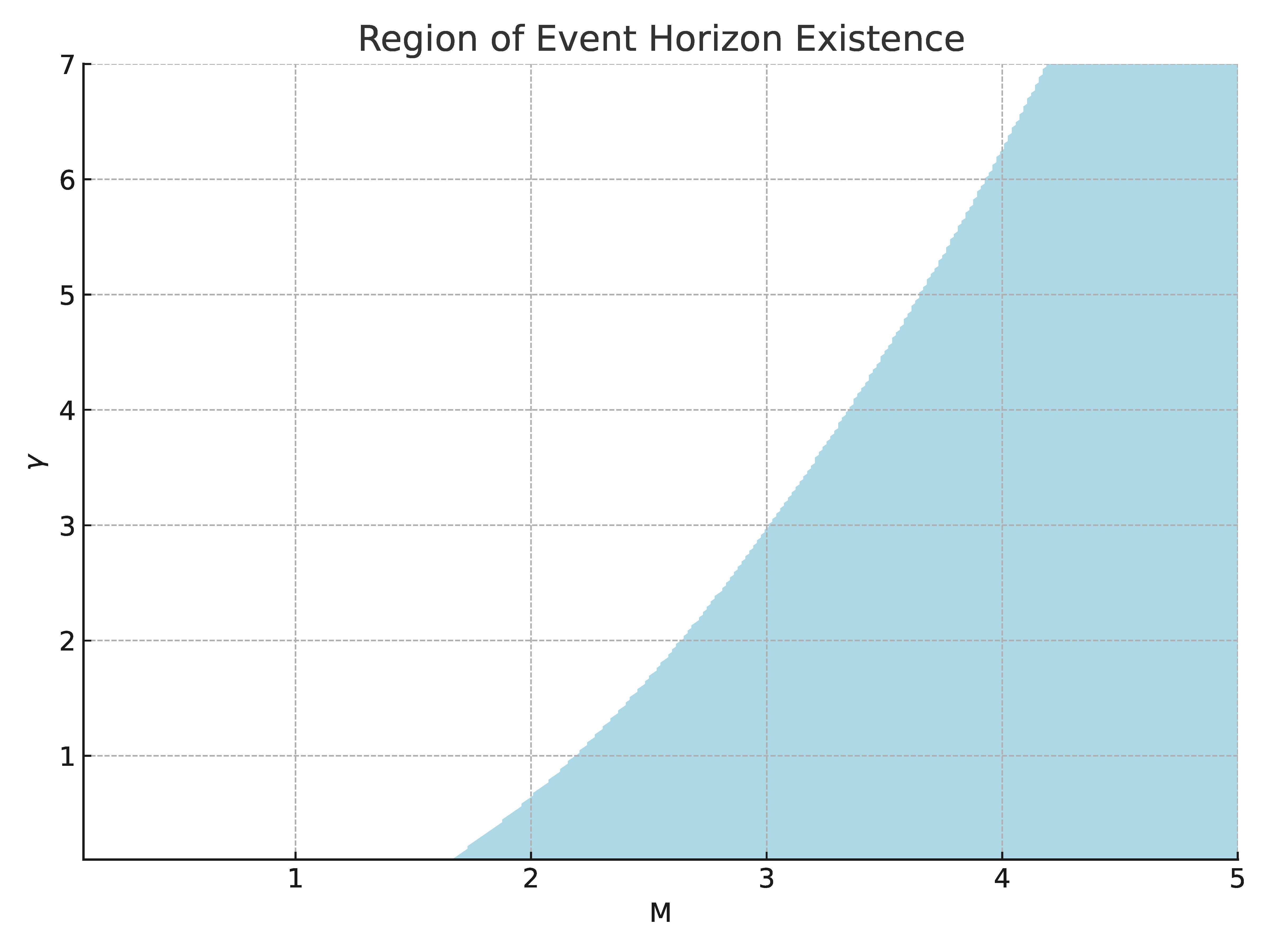}
\caption{Parametric
 range allowing for existence of the event horizon in the $M$-$\gamma$-plane.}\label{fig:Range}
\end{figure}

In this paper, we fix $\tilde{\omega} = 118/(15\pi)$, and treat $\gamma$ as a free parameter controlling the RG scale interpolation. We study the gravitational quasinormal mode spectrum of this quantum-corrected geometry and compare the results to classical expectations. The parametric range corresponding to the black hole configuration is shown in Figure \ref{fig:Range}.

\vspace{-20pt}
\section{Effective Potential for Gravitational Perturbations}
\label{sec3}

The analysis of axial gravitational perturbations in quantum-corrected black hole spacetimes derived from the Hamiltonian constraints approach is challenging, since the background metrics do not arise as exact solutions of the Einstein field equations, but rather from effective modifications. A practical workaround, adopted in~\cite{Bouhmadi-Lopez:2020oia}, is to model quantum corrections as an effective anisotropic fluid within the classical general relativity framework. This interpretation enables the use of standard perturbative techniques to derive wave-like equations governing metric~fluctuations.

In this setup, the background metric is assumed to take the form
\begin{equation}
ds^2 = -f(r)\,dt^2 + \frac{1}{g(r)}\,dr^2 + r^2(d\theta^2 + \sin^2\theta\,d\phi^2), \label{eq:metric}
\end{equation}
where $f(r)$ and $g(r)$ encode the quantum corrections to the geometry. The stress-energy tensor of the effective anisotropic fluid is given by
\begin{equation}
T_{\mu\nu} = (\rho + p_t)u_\mu u_\nu + p_t g_{\mu\nu} + (p_r - p_t)s_\mu s_\nu, \label{eq:fluid}
\end{equation}
where $\rho$ is the energy density, $p_r$ and $p_t$ are the radial and tangential pressures respectively, $u^\mu$ is the fluid four-velocity, and $s^\mu$ is the unit radial spacelike vector orthogonal to $u^\mu$. For the static background one has
\begin{equation}
u^\mu = \left(\frac{1}{\sqrt{f(r)}}, 0, 0, 0\right), \quad s^\mu = \left(0, \sqrt{g(r)}, 0, 0\right),
\end{equation}
satisfying the normalization conditions $u_\mu u^\mu = -1$, $s_\mu s^\mu = 1$, and $u_\mu s^\mu = 0$.

Focusing on axial perturbations in the Regge–Wheeler gauge~\cite{Regge:1957td}, the perturbed metric takes the form

\begin{equation}
h_{\mu\nu}^{\text{axial}} = \left[
\begin{array}{cccc}
0 & 0 & 0 & h_0(t,r)\,\sin\theta\,\partial_\theta P_\ell(\cos\theta) \\
0 & 0 & 0 & h_1(t,r)\,\sin\theta\,\partial_\theta P_\ell(\cos\theta) \\
0 & 0 & 0 & 0 \\
h_0(t,r)\,\sin\theta\,\partial_\theta P_\ell(\cos\theta) & h_1(t,r)\,\sin\theta\,\partial_\theta P_\ell(\cos\theta) & 0 & 0 \\
\end{array}
\right],
\end{equation}
where $P_\ell$ denotes the Legendre polynomial of degree $\ell$. The axial sector is insensitive to scalar-type perturbations of $\rho$, $p_r$, and $p_t$, which transform as scalars under rotations.

By following the steps outlined in \cite{Bouhmadi-Lopez:2020oia}, the Einstein equations lead to a system of equations for the perturbation functions $h_0(t,r)$ and $h_1(t,r)$. After introducing a master variable $\Psi(t,r)$ defined by
\begin{equation}
h_1(t,r) = \frac{r}{\sqrt{f(r)g(r)}}\,\Psi(t,r),
\end{equation}
and switching to the tortoise coordinate $r_*$ defined by
\begin{equation}
\frac{dr_*}{dr} = \frac{1}{\sqrt{f(r)g(r)}},
\end{equation}
one arrives at the wave-like equation
\begin{equation}
\frac{d^2\Psi}{dr_*^2} + \left[\omega^2 - V(r)\right]\Psi = 0, \label{eq:RW}
\end{equation}
with the effective potential for axial gravitational perturbations given by \cite{Bouhmadi-Lopez:2020oia,Konoplya:2024lch}
\begin{equation}
V(r) = f(r) \left[ \frac{2g(r)}{r^2} - \frac{(f(r)g(r))'}{2rf(r)} + \frac{(\ell+2)(\ell-1)}{r^2} \right], \label{eq:Vax}
\end{equation}
where a prime denotes differentiation with respect to $r$.

\begin{figure}[H]
\centering
\includegraphics[width=0.58\textwidth]{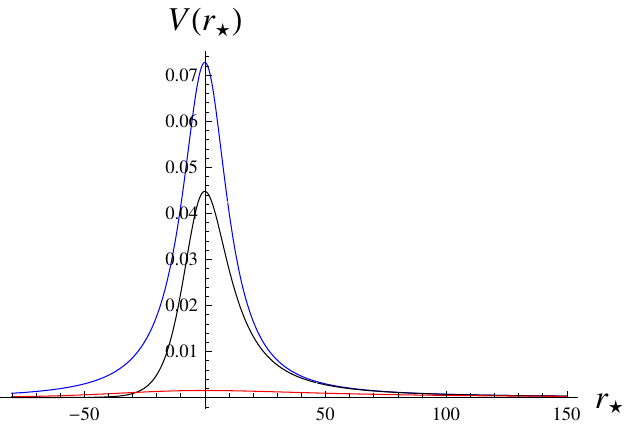}
\caption{Effective potentials for $\ell=2$ $\gamma=0.1$: $M=1.66$ (blue, top), $M=2$ (black, middle), $M=10$ (red, bottom).}\label{fig:PotentialL2}
\end{figure}

\begin{figure}[H]
\centering
\includegraphics[width=0.58\textwidth]{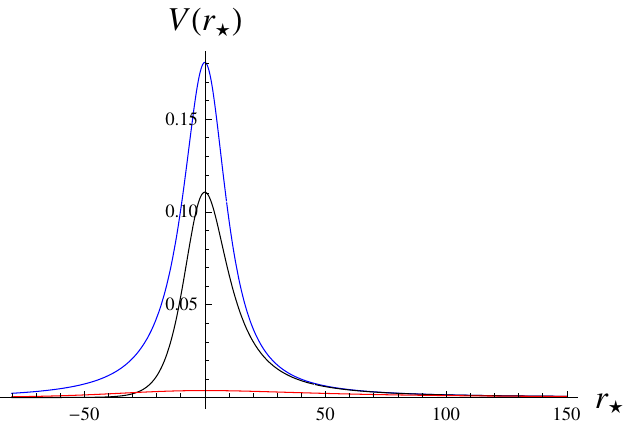}
\caption{Effective potentials for $\ell=3$ $\gamma=0.1$: $M=1.66$ (blue, top), $M=2$ (black, middle), $M=10$ (red, bottom).}\label{fig:PotentialL3}
\end{figure}

This potential reduces to the standard Regge–Wheeler potential in the Schwarzschild limit $f(r) = g(r) = 1 - 2M/r$. For small deviations from Schwarzschild geometry, this expression reliably captures the leading quantum corrections in the axial gravitational channel. The positivity of $V(r)$ (see Figures \ref{fig:PotentialL2} and \ref{fig:PotentialL3}) ensures linear stability of the perturbations under the approximations made.

\vspace{-20pt}

\vspace{-20pt}
\section{Quasinormal Modes}
\label{sec4}

Quasinormal modes (QNMs) are characteristic oscillations of perturbed black holes, governed by linear field equations subject to specific boundary conditions. They dominate the intermediate ringdown phase of a black hole’s response to perturbations and are determined entirely by the black hole’s geometry and the nature of the field. Their complex frequencies encode both the oscillation frequency (real part) and decay rate (imaginary part), with stability requiring $\text{Im}(\omega) < 0$.

\subsection{Boundary Conditions}

For spherically symmetric, asymptotically flat black hole spacetimes of the form
\begin{equation}
ds^2 = -f(r) dt^2 + \frac{dr^2}{f(r)} + r^2 d\Omega^2,
\end{equation}
the perturbation equation typically reduces to a Schrödinger-like wave equation:
\begin{equation}
\frac{d^2 \Psi}{dr_*^2} + \left[ \omega^2 - V(r) \right] \Psi = 0,
\end{equation}
where $r_*$ is the tortoise coordinate defined by $dr_*/dr = 1/f(r)$, and $V(r)$ is the effective potential depending on the spin of the perturbing field.

The quasinormal boundary conditions require purely ingoing waves at the event horizon and purely outgoing waves at spatial infinity:
\begin{equation}
\Psi(r_*) \sim 
\begin{cases}
e^{-i \omega r_*}, & r_* \to -\infty \quad (r \to r_+), \\
e^{+i \omega r_*}, & r_* \to +\infty \quad (r \to \infty).
\end{cases}
\end{equation}

\subsection{WKB Method with Padé Approximants}

\textls[-5]{One of the most effective semi-analytic methods for computing quasinormal frequencies is the Wentzel–Kramers} –Brillouin (WKB) approximation, originally developed for quantum tunneling problems and adapted for black hole perturbations in~\cite{Schutz:1985km,Iyer:1986np}. The WKB formula up to N-th order can be written as \cite{Konoplya:2003ii,Matyjasek:2017psv}:
\begin{equation}
\frac{i Q_0}{\sqrt{2 Q_0''}} - \sum_{j=2}^{N} \Lambda_j = n + \frac{1}{2}, \quad n = 0,1,2,\ldots,
\end{equation}
where $Q = \omega^2 - V(r)$ is expanded around the peak of the potential, $Q_0 = Q(r_{\max})$, $Q_0''$ is the second derivative with respect to $r_*$ at the peak, and $\Lambda_j$ are higher-order correction terms \cite{Iyer:1986np,Konoplya:2003ii,Matyjasek:2017psv}.

To improve convergence and accuracy, especially for low multipole numbers, Padé approximants were introduced in~\cite{Matyjasek:2017psv}. The WKB series is recast as a rational function:
\begin{equation}
\text{WKB-Padé}(m,p) = \frac{P_m(x)}{Q_p(x)},
\end{equation}
where the Padé approximant of order $(m,p)$ matches the WKB expansion up to order $m+p$. This technique significantly enhances precision and often yields results close to those obtained by numerical integration \cite{Bolokhov:2023ruj,Bolokhov:2024ixe,Skvortsova:2024wly,Skvortsova:2023zmj}.

\subsection{Time-Domain Integration}

An alternative and fully numerical method for extracting QNMs is time-domain integration, particularly well-suited for studying stability and late-time behavior. In this approach, the wave equation is discretized using light-cone coordinates $u = t - r_*$, $v = t + r_*$, and evolved iteratively via the characteristic integration scheme~\cite{Gundlach:1993tp}:
\begin{equation}
\Psi(N) = \Psi(W) + \Psi(E) - \Psi(S) - \frac{\Delta^2}{8} V(S) \left[ \Psi(W) + \Psi(E) \right],
\end{equation}
where $N$, $S$, $E$, and $W$ denote points on a grid and $\Delta$ is the step size. By specifying appropriate initial data (e.g., a Gaussian pulse), one can monitor the evolution of the field and extract dominant frequencies using fitting techniques such as the Prony method or filtered Fourier transform.

\subsection{Discussion of Numerical Results}

In this work, we employ both the WKB method with Padé approximants and the time-domain integration to ensure the reliability of our quasinormal mode results and to explore the behavior of perturbations in the Bonanno–Reuter black hole geometry.

The quasinormal mode spectra calculated using the sixth-order WKB method with Padé approximants reveal a consistent and smooth dependence on both the black hole mass $M$ and the quantum gravity interpolation parameter $\gamma$. For fixed $\gamma$, increasing $M$ results in both the real and imaginary parts of the fundamental quasinormal frequencies approaching their classical Schwarzschild values (see Tables \ref{tab1}--\ref{tab4}). This is consistent with the fact that for large $M$, the RG corrections become subleading, and the geometry becomes effectively classical.

\vspace{-6pt}
\begin{table}[H]
\caption{QNMs ($n=0$) of the gravitational perturbations of the Bonanno-Reuter black hole, $\gamma=0.1$, calculated using the WKB approach at the 6th and 7th order together with the difference between them.
}
\newcolumntype{C}{>{\centering\arraybackslash}X}
\begin{tabularx}{\textwidth}{CCm{4cm}<{\centering}m{4cm}<{\centering}C}
\toprule
\boldmath{$\ell$} & \boldmath{$M$}  & \textbf{WKB-6,} \boldmath{$m=3$} & \textbf{WKB-7,} \boldmath{$m=3$} &  \boldmath{$\delta (\%)$} \\
\midrule
$2$ & $1.66$ & $0.263295-0.042414 i$ & $0.263300-0.042430 i$ & $0.0061\%$\\
$2$ & $1.8$ & $0.235360-0.042547 i$ & $0.235363-0.042563 i$ & $0.0066\%$\\
$2$ & $2.$ & $0.205631-0.040400 i$ & $0.205638-0.040396 i$ & $0.0037\%$\\
$2$ & $2.5$ & $0.158207-0.033963 i$ & $0.158176-0.033962 i$ & $0.0194\%$\\
$2$ & $3.$ & $0.129398-0.028806 i$ & $0.129394-0.028810 i$ & $0.0045\%$\\
$2$ & $3.5$ & $0.109725-0.024912 i$ & $0.109728-0.024919 i$ & $0.0064\%$\\
$2$ & $4.$ & $0.095368-0.021915 i$ & $0.095373-0.021918 i$ & $0.0055\%$\\
$2$ & $5.$ & $0.075714-0.017632 i$ & $0.075717-0.017628 i$ & $0.0063\%$\\
$2$ & $10.$ & $0.037484-0.008874 i$ & $0.037484-0.008875 i$ & $0.002\%$\\
$3$ & $1.66$ & $0.420450-0.044482 i$ & $0.420450-0.044482 i$ & $0.00014\%$\\
$3$ & $1.8$ & $0.375766-0.044546 i$ & $0.375766-0.044545 i$ & $0.00007\%$\\
$3$ & $2.$ & $0.328538-0.042225 i$ & $0.328538-0.042224 i$ & $0.0002\%$\\
$3$ & $2.5$ & $0.253123-0.035433 i$ & $0.253124-0.035433 i$ & $0.0001\%$\\
$3$ & $3.$ & $0.207181-0.030050 i$ & $0.207181-0.030049 i$ & $0\%$
\\
$3$ & $3.5$ & $0.175787-0.025983 i$ & $0.175787-0.025983 i$ & $0\%$\\
$3$ & $4.$ & $0.152837-0.022851 i$ & $0.152837-0.022852 i$ & $0\%$\\
$3$ & $5.$ & $0.121384-0.018382 i$ & $0.121384-0.018382 i$ & $0\%$\\
$3$ & $10.$ & $0.060127-0.009252 i$ & $0.060127-0.009252 i$ & $0\%$\\
\bottomrule
\end{tabularx}
\label{tab1}
\end{table}

\vspace{-12pt}
\begin{table}[H]
\caption{QNMs ($n=0$) of the gravitational perturbations the Bonanno-Reuter black hole, $\gamma=1$, calculated using the WKB approach at the 6th and 7th order together with the difference between them.
}
\newcolumntype{C}{>{\centering\arraybackslash}X}
\begin{tabularx}{\textwidth}{CCm{4cm}<{\centering}m{4cm}<{\centering}C}
\toprule
\boldmath{$\ell$} & \boldmath{$M$}  & \textbf{WKB-6,} \boldmath{$m=3$} & \textbf{WKB-7,} \boldmath{$m=3$} &  \boldmath{$\delta (\%)$} \\
\midrule
$2$ & $2.19$ & $0.190298-0.033193 i$ & $0.190304-0.033216 i$ & $0.0120\%$\\
$2$ & $2.25$ & $0.183920-0.033124 i$ & $0.183928-0.033142 i$ & $0.0103\%$\\
$2$ & $2.5$ & $0.161737-0.031754 i$ & $0.161737-0.031752 i$ & $0.0009\%$\\
$2$ & $3.$ & $0.131172-0.027868 i$ & $0.131145-0.027837 i$ & $0.0309\%$\\
$2$ & $3.5$ & $0.110786-0.024389 i$ & $0.110786-0.024372 i$ & $0.0155\%$\\
$2$ & $4.$ & $0.096036-0.021585 i$ & $0.096039-0.021595 i$ & $0.0104\%$\\
$2$ & $4.5$ & $0.084844-0.019328 i$ & $0.084848-0.019335 i$ & $0.0087\%$\\
$2$ & $5.$ & $0.076034-0.017481 i$ & $0.076039-0.017485 i$ & $0.0073\%$\\
$2$ & $10.$ & $0.037522-0.008858 i$ & $0.037522-0.008858 i$ & $0.002\%$\\
$3$ & $2.19$ & $0.304323-0.034901 i$ & $0.304323-0.034901 i$ & $0\%$\\
$3$ & $2.25$ & $0.294033-0.034804 i$ & $0.294034-0.034804 i$ & $0.0002\%$\\
$3$ & $2.5$ & $0.258559-0.033262 i$ & $0.258559-0.033262 i$ & $0.00017\%$\\
$3$ & $3.$ & $0.209860-0.029103 i$ & $0.209860-0.029102 i$ & $0.0002\%$\\
$3$ & $3.5$ & $0.177334-0.025467 i$ & $0.177334-0.025467 i$ & $0\%$\\
$3$ & $4.$ & $0.153820-0.022534 i$ & $0.153820-0.022534 i$ & $0\%$\\
$3$ & $4.5$ & $0.135944-0.020168 i$ & $0.135944-0.020168 i$ & $0\%$\\
$3$ & $5.$ & $0.121858-0.018234 i$ & $0.121858-0.018234 i$ & $0.0001\%$\\
$3$ & $10.$ & $0.060182-0.009235 i$ & $0.060182-0.009235 i$ & $0\%$\\
\bottomrule
\end{tabularx}
\label{tab2}
\end{table}

\begin{table}[H]
\caption{QNMs ($n=0$) of the gravitational perturbations the Bonanno-Reuter black hole, $\gamma=9/2$, calculated using the WKB approach at the 6th and 7th order together with the difference between them.
}
\newcolumntype{C}{>{\centering\arraybackslash}X}
\begin{tabularx}{\textwidth}{CCm{4cm}<{\centering}m{4cm}<{\centering}C}
\toprule
\boldmath{$\ell$} & \boldmath{$M$}  & \textbf{WKB-6,} \boldmath{$m=3$} & \textbf{WKB-7,} \boldmath{$m=3$} &  \boldmath{$\delta (\%)$} \\
\midrule
$2$ & $3.5$ & $0.110786-0.024389 i$ & $0.110786-0.024372 i$ & $0.0155\%$\\
$2$ & $3.6$ & $0.107470-0.023775 i$ & $0.107470-0.023774 i$ & $0.0007\%$\\
$2$ & $3.7$ & $0.104354-0.023190 i$ & $0.104354-0.023196 i$ & $0.0061\%$\\
$2$ & $3.8$ & $0.101420-0.022631 i$ & $0.101421-0.022640 i$ & $0.0091\%$\\
$2$ & $3.9$ & $0.098652-0.022097 i$ & $0.098654-0.022107 i$ & $0.0102\%$\\
$2$ & $4.$ & $0.096036-0.021585 i$ & $0.096039-0.021595 i$ & $0.0104\%$\\
$2$ & $4.5$ & $0.084844-0.019328 i$ & $0.084848-0.019335 i$ & $0.0087\%$\\
$2$ & $5.$ & $0.076034-0.017481 i$ & $0.076039-0.017485 i$ & $0.0073\%$\\
$2$ & $10.$ & $0.037522-0.008858 i$ & $0.037522-0.008858 i$ & $0.002\%$\\
$3$ & $3.5$ & $0.177334-0.025467 i$ & $0.177334-0.025467 i$ & $0\%$\\
$3$ & $3.6$ & $0.172051-0.024828 i$ & $0.172051-0.024828 i$ & $0\%$\\
$3$ & $3.7$ & $0.167085-0.024216 i$ & $0.167085-0.024216 i$ & $0\%$\\
$3$ & $3.8$ & $0.162408-0.023631 i$ & $0.162408-0.023631 i$ & $0\%$\\
$3$ & $3.9$ & $0.157993-0.023071 i$ & $0.157993-0.023071 i$ & $0\%$\\
$3$ & $4.$ & $0.153820-0.022534 i$ & $0.153820-0.022534 i$ & $0\%$\\
$3$ & $4.5$ & $0.135944-0.020168 i$ & $0.135944-0.020168 i$ & $0\%$\\
$3$ & $5.$ & $0.121858-0.018234 i$ & $0.121858-0.018234 i$ & $0.0001\%$\\
$3$ & $10.$ & $0.060182-0.009235 i$ & $0.060182-0.009235 i$ & $0\%$\\
\bottomrule
\end{tabularx}
\label{tab3}
\end{table}

\vspace{-12pt}
\begin{table}[H]
\caption{QNMs ($n=1$) of the gravitational perturbations the Bonanno-Reuter black hole, $\gamma=9/2$, calculated using the WKB approach at the 6th and 7th order together with the difference between them.
}
\newcolumntype{C}{>{\centering\arraybackslash}X}
\begin{tabularx}{\textwidth}{CCm{4cm}<{\centering}m{4cm}<{\centering}C}
\toprule
\boldmath{$\ell$} & \boldmath{$M$}  & \textbf{WKB-6,} \boldmath{$m=3$} & \textbf{WKB-7,} \boldmath{$m=3$} &  \boldmath{$\delta (\%)$} \\
\midrule
$2$ & $3.5$ & $0.104567-0.074770 i$ & $0.104746-0.074581 i$ & $0.202\%$\\
$2$ & $3.6$ & $0.101353-0.072894 i$ & $0.101432-0.072777 i$ & $0.113\%$\\
$2$ & $3.7$ & $0.098333-0.071100 i$ & $0.098363-0.071040 i$ & $0.0557\%$\\
$2$ & $3.8$ & $0.095491-0.069386 i$ & $0.095497-0.069368 i$ & $0.0158\%$\\
$2$ & $3.9$ & $0.092812-0.067746 i$ & $0.092808-0.067761 i$ & $0.0136\%$\\
$2$ & $4.$ & $0.090283-0.066178 i$ & $0.090277-0.066217 i$ & $0.0360\%$\\
$2$ & $4.5$ & $0.079507-0.059269 i$ & $0.079521-0.059362 i$ & $0.0952\%$\\
$2$ & $5.$ & $0.071086-0.053622 i$ & $0.071125-0.053720 i$ & $0.118\%$\\
$2$ & $10.$ & $0.034827-0.027228 i$ & $0.034863-0.027232 i$ & $0.082\%$\\
$3$ & $3.5$ & $0.173288-0.077127 i$ & $0.173290-0.077120 i$ & $0.0039\%$\\
$3$ & $3.6$ & $0.168080-0.075202 i$ & $0.168082-0.075195 i$ & $0.0037\%$\\
$3$ & $3.7$ & $0.163188-0.073358 i$ & $0.163189-0.073352 i$ & $0.0035\%$\\
$3$ & $3.8$ & $0.158582-0.071593 i$ & $0.158583-0.071587 i$ & $0.0033\%$\\
$3$ & $3.9$ & $0.154238-0.069902 i$ & $0.154239-0.069897 i$ & $0.0031\%$\\
$3$ & $4.$ & $0.150133-0.068283 i$ & $0.150134-0.068278 i$ & $0.0029\%$\\
$3$ & $4.5$ & $0.132573-0.061134 i$ & $0.132573-0.061131 i$ & $0.0020\%$\\
$3$ & $5.$ & $0.118763-0.055283 i$ & $0.118763-0.055282 i$ & $0.0013\%$\\
$3$ & $10.$ & $0.058535-0.028017 i$ & $0.058535-0.028018 i$ & $0.0008\%$\\
\bottomrule
\end{tabularx}
\label{tab4}
\end{table}

\begin{figure}[H]
\centering
\includegraphics[scale=0.6]{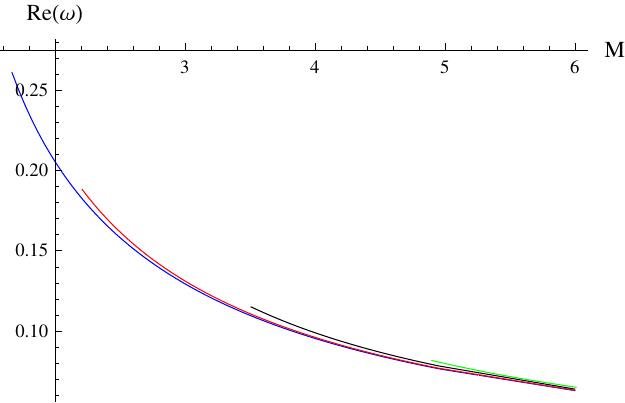}
\includegraphics[scale=0.6]{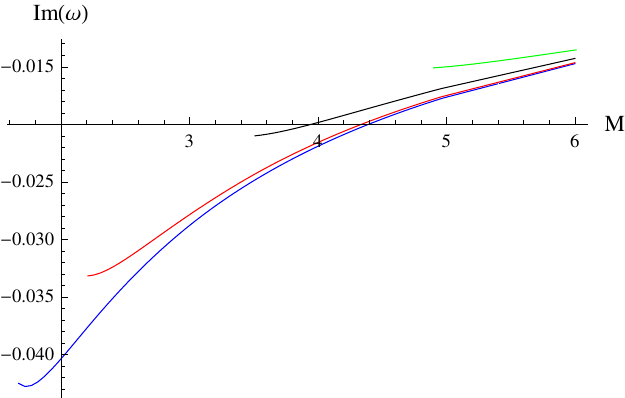}
\caption{Real and imaginary part of the dominant ($n=0$) QNMs of the $\ell=2$ gravitational perturbations for the Bonanno-Reuter black hole calculated by WKB6Pade3 as functions of $M$,  $\gamma=0.01$ (blue), $\gamma=1$ (red), $\gamma=9/2$ (black), $\gamma=10$ (green).}\label{qnmsL2}
\end{figure}

\begin{figure}[H]
\centering
{\includegraphics[scale=0.6]{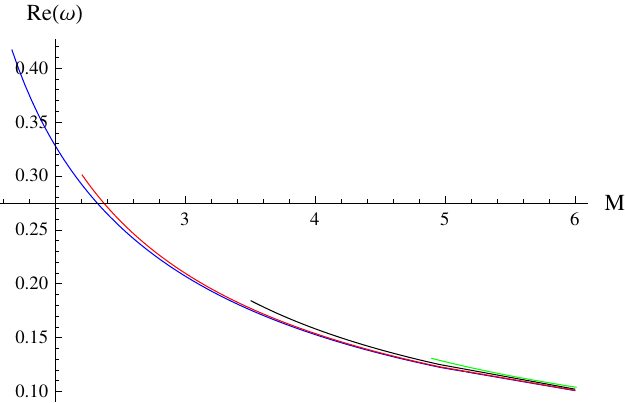}
\includegraphics[scale=0.6]{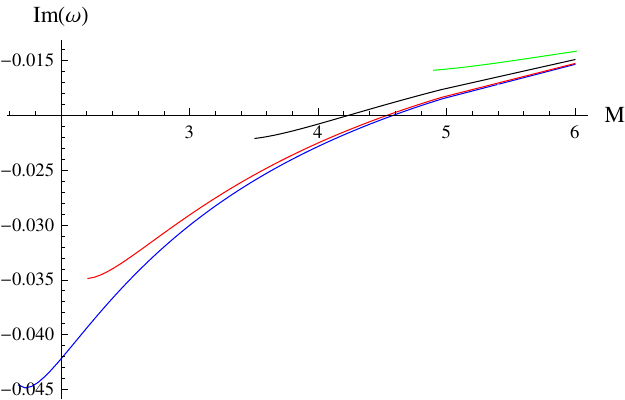}}
\caption{Real and imaginary part of the dominant ($n=0$) QNMs of the $\ell=3$ gravitational perturbations for the Bonanno-Reuter black hole calculated by WKB6Pade3 as functions of $M$,  $\gamma=0.01$ (blue), $\gamma=1$ (red), $\gamma=9/2$ (black), $\gamma=10$ (green).}\label{qnmsL3}
\end{figure}

Similarly, for fixed $M$, increasing $\gamma$ tends to suppress the deviation from the classical limit, reinforcing the interpretation of $\gamma$ as controlling the scale at which quantum effects become significant. In particular, values of $\gamma \gtrsim 5$ already lead to frequencies indistinguishable (within numerical error) from Schwarzschild for $M \gtrsim 5$. This trend supports the idea that the Bonanno--Reuter black hole smoothly interpolates between quantum-corrected and classical behavior depending on the choice of parameters. As shown in Figures 
\ref{qnmsL2} and \ref{qnmsL3}, the quantum corrected black hole of the same mass as its classical counterpart has longer lived modes with slightly smaller oscillations rate.

\begin{figure}[H]
\centering
\includegraphics{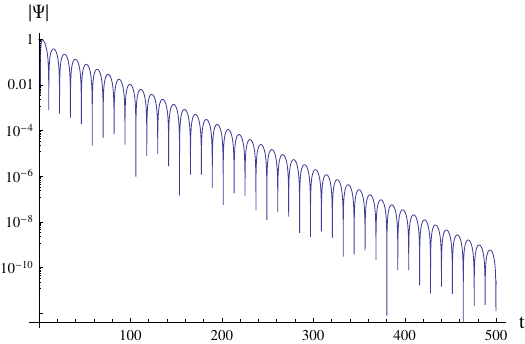}
\caption{Time-domain profile for $\ell=2$ $\gamma=0.1$: $M=1.66$. The Prony method gives $\omega= 0.263305 - 0.042429 i$, while the WKB data is $\omega = 0.263295 - 0.042414 i$. }
\label{fig6}
\end{figure}

\vspace{-12pt}
The WKB method with Padé approximants of order $m=3$ produces consistent results across different multipoles and parameter values. As seen in Tables \ref{tab1}--\ref{tab4}, the relative errors between sixth- and seventh-order WKB calculations are typically $\delta \lesssim 0.01\%$, which is one to two orders of magnitude smaller than the deviations from Schwarzschild frequencies. This confirms the robustness of the results and ensures that the observed deviations are genuine physical effects rather than numerical artifacts. As shown in Figure \ref{fig6},
the frequency extracted from the time-domain profile using the Prony method matches the sixth-order WKB result to within <0.005\%, reinforcing the reliability of the WKB approach for moderate $\ell$.

\vspace{-20pt}
\section{Grey-Body Factors}
\label{sec5}

In the semiclassical picture of Hawking radiation, black holes emit particles with a thermal spectrum at a temperature determined by the surface gravity at the horizon. However, the actual radiation observed at infinity deviates from perfect blackbody behavior due to the backscattering of waves by the spacetime curvature outside the horizon. This leads to the introduction of \emph{grey-body factors}, which act as frequency-dependent transmission coefficients modulating the emitted radiation spectrum.

Unlike QNMs, for which both boundaries are radiative (ingoing at the horizon and outgoing at infinity), the setup for grey-body factors corresponds to a scattering problem:
\begin{equation}
\Psi(r_*) \sim 
\begin{cases}
\mathcal{T} e^{-i \omega r_*}, & r_* \to -\infty \quad (r \to r_+), \\
e^{-i \omega r_*} + \mathcal{R} e^{i \omega r_*}, & r_* \to +\infty \quad (r \to \infty),
\end{cases}
\end{equation}
where $\mathcal{R}$ and $\mathcal{T}$ are the reflection and transmission coefficients, respectively. The \emph{grey-body factor} is defined as the modulus squared of the transmission coefficient:
\begin{equation}
\Gamma_\ell(\omega) = |\mathcal{T}|^2,
\end{equation}
which physically represents the probability that a wave of frequency $\omega$ and angular momentum $\ell$ escapes to infinity.

Grey-body factors can be computed semi-analytically using the WKB method~\cite{Schutz:1985km,Iyer:1986np,Konoplya:2019hlu}, especially in the regime where the effective potential has a single well-defined barrier. The WKB transmission coefficient at a given order yields:
\begin{equation}
\Gamma_\ell(\omega) = \left( 1 + e^{2 \pi K} \right)^{-1},
\end{equation}
where $K$ is a WKB quantity defined by:
\begin{equation}
K = \frac{i (\omega^2 - V_0)}{\sqrt{-2 V_0''}} + \cdots,
\end{equation}
with $V_0$ and $V_0''$ being the value and second derivative of the effective potential at its maximum, respectively. Higher-order corrections can be included, and as in the case of QNMs, Padé resummation should improve convergence and numerical accuracy~\cite{Matyjasek:2017psv}.
This method is particularly reliable for frequencies near the peak of the potential, and especially when $\ell$ is large. Here we  used the 6th order WKB formula \cite{Konoplya:2003ii} as usually it provides the best accuracy.

An alternative and increasingly explored approach relies on the recently proposed correspondence between grey-body factors and the fundamental quasinormal mode in the eikonal regime~\cite{Konoplya:2024lir}. In this framework, the grey-body factor is approximately expressed as:
\begin{equation}
\Gamma_\ell(\omega) \approx \left[ 1 + \exp\left( \frac{2 \pi (\omega^2 - \text{Re}[\omega_0]^2)}{4\, \text{Re}[\omega_0]\, |\text{Im}[\omega_0]|} \right) \right]^{-1} + \mathcal{O}(\ell^{-1}),
\end{equation}
where $\omega_0$ is the least damped quasinormal frequency for a given multipole number $\ell$. This approximation has been shown to be remarkably accurate for large $\ell$, and even for moderate values when higher-order corrections in the overtone number are included \cite{Bolokhov:2024otn,Skvortsova:2024msa,Dubinsky:2024vbn,Malik:2024cgb,Lutfuoglu:2025ohb,Lutfuoglu:2025ljm,Lutfuoglu:2025hjy}.

In this work, we compute the grey-body factors for gravitational perturbations in the Bonanno–Reuter black hole background using both methods. Their agreement offers a valuable cross-check and enhances our understanding of wave propagation in quantum-corrected spacetimes.

The behavior of grey-body factors in the Bonanno--Reuter geometry mirrors the features of the effective potential, particularly its height and width, which influence the transmission probability of gravitational waves through the curvature barrier. Although here the full profile of the potential is only presented for varying mass at fixed $\gamma$, it can be inferred that increasing $M$ lowers and widens the potential peak, thereby enhancing the transmission probability and raising the grey-body factors for a fixed frequency $\omega$.

\begin{figure}[H]
\centering
\includegraphics[scale=0.9]{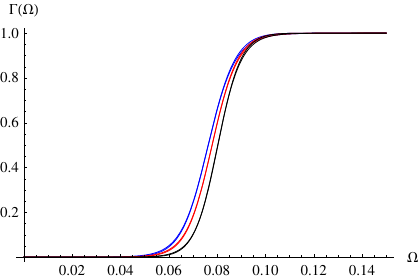}
\includegraphics[scale=0.9]{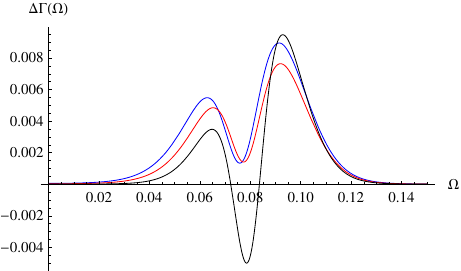}
\caption{Grey-body factors for $\ell=2$ case calculated via the 6th order method and with the help of the correspondence with QNMs (left) and the difference between the results obtained by the two methods (right): $\gamma=0.01$ (blue), $\gamma=9/2$ (red), $\gamma=10$ (black).}\label{fig:GBFL2}
\end{figure}

\begin{figure}[H]
\centering
\includegraphics[scale=0.9]{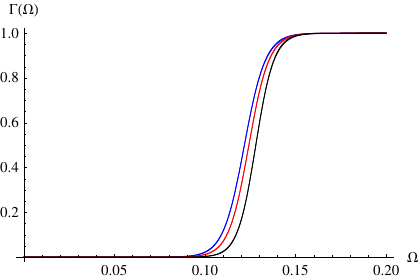}
\includegraphics[scale=0.9]{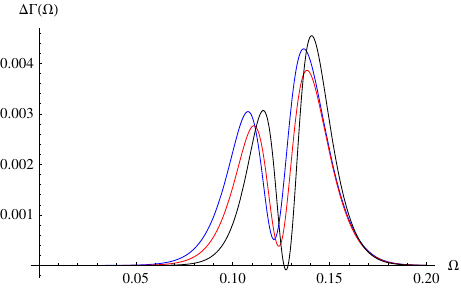}
\caption{Grey-body factors for $\ell=3$ case calculated via the 6th order method and with the help of the correspondence with QNMs (left) and the difference between the results obtained by the two methods (right): $\gamma=0.01$ (blue), $\gamma=9/2$ (red), $\gamma=10$ (black).}\label{fig:GBFL3}
\end{figure}

This interpretation is consistent with the computed grey-body factors shown in Figures \ref{fig:GBFL2} and \ref{fig:GBFL3}.
At fixed $\ell$, increasing $M$ generally leads to a higher transmission coefficient. The impact of varying $\gamma$ is somewhat subtler but can be traced to the way it affects the near-horizon structure of the metric and thus shifts the effective potential. Notably, the dependence of grey-body factors on $\gamma$ is more pronounced at low $M$, where quantum corrections are strongest. When mass is fixed, larger $\gamma$ corresponds to higher effective potential and smaller grey-body factors, as shown in Figures \ref{fig:GBFL2} and \ref{fig:GBFL3}. 

\vspace{-12pt}
\vspace{-26pt}
\vspace{-12pt}

Moreover, the comparison between grey-body factors obtained via the WKB method and those predicted by the QNM--grey-body correspondence (equation~(23)) confirms the reliability of both methods. The agreement is excellent for $\ell = 3$, and even for $\ell = 2$, the discrepancy is below 1\% across most frequencies, validating the use of the eikonal-mode approximation even in moderately low $\ell$ regimes. These results provide additional evidence for the utility of QNMs as probes of black hole spacetimes beyond classical general relativity.

\section{Conclusions}
\label{sec6}

In this work, we have computed and analyzed the gravitational QNMs and grey-body factors of the Bonanno--Reuter regular black hole---a quantum-corrected geometry emerging from the asymptotic safety program. By fixing the RG parameter $\tilde{\omega}$ and varying the interpolation parameter $\gamma$, we have explored how quantum modifications affect the observable signatures of black holes.

Our results demonstrate that both the real and imaginary parts of quasinormal frequencies smoothly converge to their classical Schwarzschild limits for large masses or large values of $\gamma$. This confirms that the classical regime is recovered in the appropriate limit and validates the consistency of the RG-improved metric. Furthermore, the calculated grey-body factors exhibit a corresponding behavior, with quantum corrections modifying the transmission spectrum in a predictable manner based on the deformation of the effective potential. The quantum corrected black holes of the same mass as the Schwarzschild one have longer lived QNMs and smaller grey-body factors. 

Importantly, the accuracy of the WKB method with Padé resummation is sufficient to resolve even subtle quantum-induced deviations from the Schwarzschild case, and the consistency with QNM--grey-body correspondence further strengthens the reliability of the findings.

It is worth noting that QNMs and grey-body factors have also been studied for several other black hole models within the framework of Asymptotically Safe Gravity~\cite{Konoplya:2023aph,Lutfuoglu:2025ohb,Zinhailo:2023xdz,Stashko:2024wuq}.

Altogether, our study provides a systematic understanding of how quantum gravity corrections within asymptotic safety may influence gravitational wave observables. This offers a valuable benchmark for future studies involving time-domain evolution, stability analysis, or potential observational constraints.

%	\section*{Supplementary Materials} 
% The following supporting information can be downloaded at: URL, Figure S1: title; Table S1: title; Video S1: title.
 
		\section*{Author Contributions}
S.B.: conceptualization, methodology, software, writing—original draft preparation; M.S.: methodology, software, visualization, writing—reviewing and editing. All authors have read and agreed to the published version of the manuscript.

		\section*{Funding}

This research was funded by RUDN University, grant number FSSF-2023-0003.

		\section*{Data Availability Statement}

Not applicable.

		\section*{Conflicts of Interest}
The authors declare no conflict of interest.

\end{document}